\documentclass[twocolumn,groupedaddress,amsmath,aps,pra]{revtex4-1}
\usepackage{graphicx}

\begin{document}
\title{Radiation trapping and L\'{e}vy flights in atomic vapours: an introductory review}
\author{Martine Chevrollier}
\affiliation{Departamento de F\'{i}sica, Universidade Federal da Para\'{i}ba\\
                 Jo\~{a}o Pessoa, PB, Brazil\\
        }

\begin{abstract}
Multiple scattering is a process in which a particle is repeatedly deflected by other particles. In an overwhelming majority of cases, the ensuing random walk can successfully be described through Gaussian, or normal, statistics. However, like a (growing) number of other apparently inofensive systems, diffusion of light in dilute atomic vapours eludes this familiar interpretation, exhibiting a superdiffusive behavior. As opposed to normal diffusion, whereby the particle executes steps in random directions but with lengths slightly varying around an average value (like a drunkard whose next move is unpredictable but certain to within a few tens of centimeters), superdiffusion is characterized by sudden abnormally long steps (L\'{e}vy flights) interrupting sequences of apparently regular jumps which, although very rare, determine the whole dynamics of the system. The formal statistics tools to describe superdiffusion already exist and rely on stable, well understood distributions. As scientists become aware of, and more familiar with, this non-orthodox possibility of interpretation of random phenomena, new systems are discovered or re-interpreted as following L\'{e}vy statistics. Propagation of light in resonant atomic vapours is one of these systems that have been studied for decades and have only recently been shown to be the scene of L\'{e}vy flights.
\end{abstract}

\maketitle

\section{Introduction}

As a branch of non-equilibrium statistical physics, the study of transport phenomena seeks information on \textit{microscopic processes} through measurement of \textit{macroscopic physical quantities} associated with transport, such as a diffusion coefficient, generally through mean square displacements. An emblematic diffusion example is still the case of particles motion in a ``granular" medium. This problem of so-called Brownian motion was explained by Einstein \cite{Einstein1905}. This is a typical case of normal statistics system and seems to be obeyed by an enormous number of examples, where displacements have arbitrary values but where very precise average values characterize the system. Examples can be found in all kinds of scientific fields, from economics and finances, biology and chemistry to astronomy and physics. The interpretation of these stochastic processes is done using well known tools from Statistics.
These are basically grounded in two main laws: i) the law of large numbers, stating that as you increase the number of trials the resulting averages and moments tend to the theoretical ones and ii) the Central Limit Theorem (CLT), which states that the accumulated action of a large number, $n$, of equivalent (same probability distribution, with expected value $\mu$ and variance $\sigma^2$), independent individual random processes results in a Gaussian distribution with expected value and variance fully determined by those ($\mu$ and $\sigma^2$) of the generating distribution, no matter the \textit{shape} of this generating distribution. This insensitivity to the details of the microscopic process explains why many physical, chemical, social, financial and so on phenomena lend themselves to this interpretation, so that it is commonly called ``normal". 
These normal laws applied to diffusion phenomena lead to a mean square displacement varying linearly with time, i.e. after a large number of random kicks in all directions, a particle of coffee in a cup of milk (or molecules of oxygen in the air, or a drunkard with no purpose) will have covered a distance proportional to the square root of the time spent: 
\begin{equation}
\left\langle r^2 \right\rangle\propto t.
\end{equation}
The first transport phenomenon breaking this law that comes to mind is the ballistic movement where a particle, not suffering any collision (and thus actually non diffusive), travels in a straight line with a speed $v$, covering during a time $t$ the distance 
\begin{equation}
r=vt, 
\end{equation}

\noindent i.e. 

\begin{equation}
r^2 \propto t^2.
\label{ballistic}
\end{equation}

\noindent Diffusion is actually ballistic as long as the observation time is shorter than the mean time between collisions. Other transport phenomena break the normal laws more radically and one of the most recently observed is the object of this paper: the superdiffusion of light in resonant atomic vapours. The diffusion problem we are discussing here is related to the \textit{dynamical} behavior of the system. This dynamical problem has a ``long tail" signature, a characteristic it shares with another class of scale-free systems, the web and equivalent systems based on connectivity. The peculiar structure or topology of those systems leads to statistical distributions following power laws, resulting in unexpected behaviors such as the dominant role of a very few elements of the system \cite{Brockmann2003}. Let us therefore distinguish these categories from the dynamical behavior of systems being discussed here: a dynamical system may, for example, \textit{evolve} from normal to sub- or super-diffusive. Focussing on the particular case of light diffusion, we will introduce here the fascinating concept of L\'{e}vy Flights, where the 'average' response may become meaningless, because the concepts of mean value or variance are no longer well-defined.

\section{Normal statistics}
In order to better appreciate how abnormal superdiffusion is, let us here summarize the main results of Normal (Gaussian) Statistics: Let $x$ be a set of identically distributed, independent variables $x_i$, with probability density $p(x)$, expected value $\left\langle x\right\rangle=\mu$ and variance $\sigma^2=\left\langle x^2\right\rangle-\left\langle x\right\rangle^2$. According to the law of large numbers, a random sampling of the distribution $p(x)$ will yield a sample sum 

\begin{equation}
X_n=\sum_{i=1}^{n}{x_i},\label{sum}
\end{equation}

\noindent a mean value

\begin{eqnarray}
\mu_n=\frac{1}{n}X_n,\label{mean}\\
\textrm{ with } \lim_{n \to +\infty}\mu_n = \mu ,\nonumber
\end{eqnarray}

\noindent and a variance
 
\begin{eqnarray}
\sigma_n^{2}=\frac{1}{n}\sum_{i=1}^{n}{(x_i-\mu)^{2}},\\
\textrm{ with } \lim_{n \to +\infty}\sigma_n^{2} = \sigma^{2} .\nonumber
\end{eqnarray}

\noindent A very interesting result of this statistical treatment of random events is that, regardless of the shape of most distributions $p(x)$, the distribution of sums, $p(X_n)$ or of mean values, $p(\mu_n)$, of random samples, tend to be normal, with limiting Gaussian distributions fully determined by the single event mean $\mu=\left\langle x\right\rangle$ and the second moment $\left\langle x^2\right\rangle$, assuming they are finite. This is the deep, powerful statement of the CLT. 

\begin{figure}
\centering
\includegraphics[width=8.0cm]{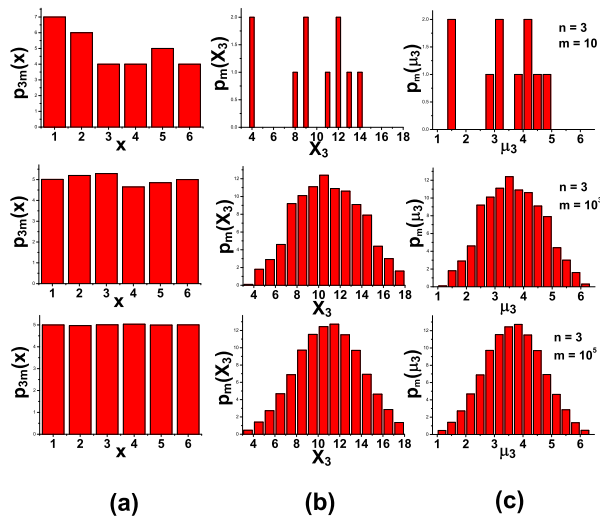}
\caption{(a) Distribution $p_nm(x)$ of $m \times n$ single events $x$ drawn from the distribution $p(x)=1/6$, 1$\leq X \leq$6 (dice rolls). The distributions (b) $p_m(X_n)$ of $m$ samples of sums $X_n$ and (c) $p_m(\mu_n)$ of $m$ samples of mean values $\mu_n$ tend to Gaussian distributions as $m$ is increased, with mean and variance determined by expected value and variance of the parent distribution $p(x)$ (see text). For $n=3$ and: top, $m=10$; middle $m=10^3$; bottom $m=10^5$.}
\label{Distributionsa}
\end{figure}

\begin{figure}
\centering
\includegraphics[width=8.0cm]{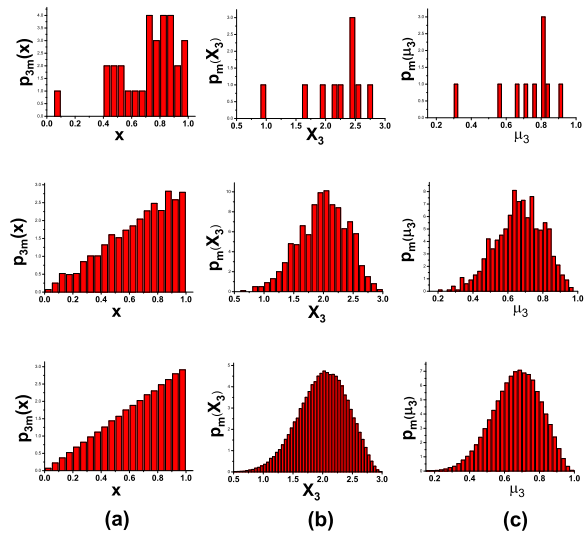}
\caption{Same as figure \ref{Distributionsa}, with $p(x) = x$,\hspace{3 mm} 0$\leq X \leq$1.}
\label{Distributionsb}
\end{figure}

In Figs. \ref{Distributionsa} and \ref{Distributionsb} are shown two examples of non-Gaussian distributions of variables whose sums and averages tend to be normally distributed. The first one (figure \ref{Distributionsa}), an example of a discrete uniform distribution, is the distribution of probability of rolling any value of a fair 6-sided die in a single roll. The second one (figure \ref{Distributionsb}) is an example of an asymmetric continuous distribution. In both examples, the sum and the mean value of $n$ draws are given by equations (\ref{sum}) and (\ref{mean}), respectively. $p_m(X_n)$ and $p_m(\mu_n)$ are the observed distributions of $m$ samples of $X_n$ and $\mu_n$, respectively. In figure (\ref{Distributionsa}) and (\ref{Distributionsb}) we can therefore see the CLT at work: the means of the sample sum $X_n$ and sample mean $\mu_n$ tend to approach the expected values $\left\langle X_n \right\rangle = n \mu$ and $\left\langle \mu_n \right\rangle = \mu$, while their variance tend to the theoretical variances Var$(X_n) = n \sigma^2$ and
Var$(\mu_n) = \sigma^2 / n$, as the number $m$ of samples grows. 
  
This seemingly systematic convergence explains why so many real phenomena resulting from the action of many tiny random events with very diverse distributions can eventually be fitted by a Gaussian distribution. 

\begin{figure}
\centering
\includegraphics[width=8.0cm]{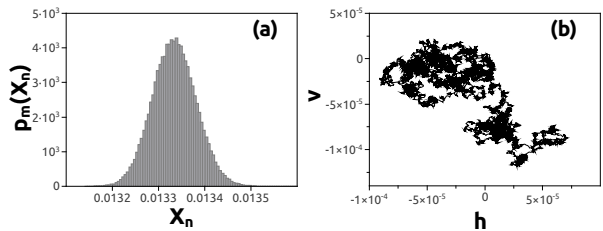}
\caption{Characterizing normal random walk in two dimensions (2D): A particle executes random walks of $n = 10^4$ steps of length $x$ drawn from the probability distribution $p(x)=1/x^5$. (a) Distribution $p_m(X_n)$ of $m$ samples of sums $X_n$ (equation (\ref{sum})) for $m = 10^5$. (b) 2D representation of one of these random walks. Each of the $n$ steps of length $x$ occurs in a random direction to a new position $(h,v)$. The initial position is (0,0). In this specific plot, the sum of step lengths is $X_n= 0.013342$ and the maximum step length is about 10 times larger than the mean one.}
\label{NormalDiff}
\end{figure}

An additional example of non-normal distribution $p(x)$ leading to normally  distributed sums $X_n$ and mean-values $\mu_n$ is shown in figure \ref{NormalDiff} for $p(x) = 1/x^5$. The 2-dimensional (2D) isotropic random walk resulting from step lengths drawn from $p(x)= 1/x^ 5$ is shown in figure \ref{NormalDiff}(b) as a visual confirmation of a diffusive, Brownian-like behavior. Assuming constant velocity, the mean-square displacement $\left\langle X^2\right\rangle =D t^{\gamma} \propto t^{\gamma}$, with $\gamma =1$. Let us now generalize these findings by focussing on probability distributions whose \textit{asymptotic decay} can be approximated by a power-law:

\begin{eqnarray}
p(x)\longrightarrow \frac{1}{x^{\alpha}} \label{asympt}\\
x\longrightarrow \infty \nonumber,
\end{eqnarray}

\noindent with $\alpha >1$.

If $\alpha >3$, the first and second moments of the distribution, $\left\langle x\right\rangle$ and $\left\langle x^2\right\rangle$, are finite and the distribution obeys the CLT. This is what we observed for $\alpha=5$ in the example given in figure \ref{NormalDiff}. However, if the distribution's asymptotic decay is \textit{slower} than $1/x^3$, i.e. if $1< \alpha \leq 3$, then $\left\langle x^2\right\rangle$ is \textit{no longer finite} and the CLT does not apply anymore. In the case $1< \alpha \leq 2$, even the first moment $\left\langle x\right\rangle$ diverges, meaning that these distributions do not exhibit a typical behavior, because the probability of very large values of $x$ (far wings of the probability distribution) is much-larger-than-normal. Their repeated application (large number $n$ of events $x_i$), characterized by $X_n$ defined in equation (\ref{sum}), does not follow a Gaussian distribution. These distributions obey, however, a Generalised Central Limit Theorem, according to which, if the distribution's asymptotic behavior follows equation (\ref{asympt}) with $0< \alpha \leq 3$, then the normalized probability densities of its sum $X_n$ and, consequently, of its mean value $\mu_n$, follow a stable L\'{e}vy law.

\begin{figure}
\centering
\includegraphics[width=8.0cm]{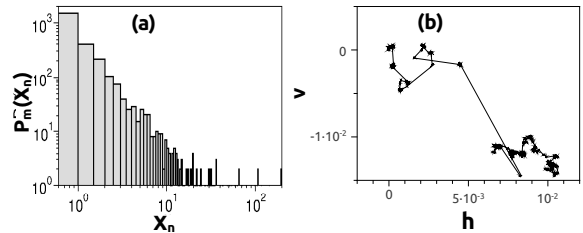}
\caption{Same as figure \ref{NormalDiff}, for $p(x)=1/x^2$. Notice the log-log scales in (a). In the specific plot in (b), the sum of the $n = 10^4$ step lengths is $X_n=0.0986$, only approximately 7 times larger than the maximum step length, which is itself about 1500 times larger than the mean one: a few particularly long steps dominate this superdiffusive random walk.}
\label{SuperDiff}
\end{figure}

In figure \ref{SuperDiff} are shown the same features as in figure \ref{NormalDiff}, this time for a superdiffusive process, namely for the probability distribution of single steps $f(x)=1/x^2$ ($\alpha = 2$). The distributions $p(X_n)$ and $p(\mu_n)$ (not shown) \textit{do not} converge to a Gaussian shape and the 2D isotropic random walk resulting from drawing the step lengths from $p(x)= 1/x^2$ does not look diffusive. Instead, we can observe in figure \ref{SuperDiff}(b) long jumps characteristic of a L\'{e}vy flight behavior. In this case, the mean-square displacement $\left\langle x^2\right\rangle =D t^{\gamma} \propto t^{\gamma}$, with $\gamma >1$.

Systems expanding as $\left\langle x^2\right\rangle\propto t^{\gamma}$ under the action of stochastic processes are said to exhibit normal diffusion if $\gamma=1$ and superdiffusion (subdiffusion) if $\gamma>1$ ($\gamma<1$). Brownian motion is a paradigm of normal diffusion and an overwhelming majority of stochastic phenomena in all fields exhibits this behavior. Much less numerous but being gradually uncovered are examples of systems showing abnormal diffusion. They are found in very diverse systems \cite{Stanley1996} such as physiology \cite{West1994,Dieterich2008}, ecology \cite{Viswanathan1999,Bartumeus2002,Nathan2002,Ramos2004,Bartumeus2005,Viswanathan2008,Humphries2010,Viswanathan2011}, human mobility and virus propagation \cite{Brockmann2006,Gonzalez2008,Stollenwerk2009}, finance and economy \cite{Mandelbrot1960}, physics \cite{Solomon1993,Bardou1994,Venkataramani1997,Shimizu2001,Tamura2002,Drysdale2004,Sharma2006,Barthelemy2008}, astrophysics \cite{Boldyrev2003}, etc. While dynamical effects of superdiffusion can be observed and characterized in those systems, the fundamental mechanisms leading to L\'{e}vy flights are in general not elucidated and remain the subject of intense study. For instance, broad distributions may appear as a consequence of the fundamental non linear dynamics of the system \cite{Klafter1996}. In ecology, L\'{e}vy flights seem to be associated to (evolutionary) optimization of search patterns in situations of scarcity (food, mates), while Brownian movements dominate in abundance contexts \cite{Humphries2010}. As we will see later on, the physical origin of L\'{e}vy flights of photons in resonant atomic vapours lies in the specific shapes of absorption and emission spectral distributions, which themselves arise from fundamental and kinetic properties of atoms.

\section{Photons, particles of light}
From now on, we will restrict ourselves to the description of the propagation of photons in material media, i.e. we restrict the discussion to particles of light. Propagation of light in scattering media represents \textit{per se} a unique and interesting system, because it is a very familiar and ubiquitous phenomenon, yet with huge practical interest, for example as our main source of knowledge about interstellar medium or in modern photonics applications. It has now gained an extra charm as a system likely to exhibit abnormal diffusion properties and, probably, the earliest recognized as such \cite{Kenty1932}. 
 
Light represents a versatile tool for studying transport phenomena and can be analysed in a variety of materials. In free space, a photon travels in a straight line with speed $c$ and thus covers a distance $\rho=ct$ during a time interval $t$, which is a ballistic movement according to equation (\ref{ballistic}). In a (dilute) material medium, the photons are scattered by material constituents and a transition occurs from ballistic to diffusive behavior as the constituents' density increases. The mean free path is the mean distance traveled by the photons between two scattering events, $\bar{\ell}=\langle \rho \rangle=1/n \displaystyle \sum_{i=0}^{n}{\rho_i}$ (in photon scattering, dominant delays are usually the collisions times, much larger than the propagation time between two scattering events, as opposed to scattering of material particles, where the dominant time is the propagation one). If the typical dimensions of the medium are smaller than $\bar{\ell}$, the transport is still essentially ballistic, otherwise it is diffusive. In a \textit{normal} diffusive medium, the propagation of photons  can be described as a random walk with a Gaussian distribution $p(\rho)$ of the step length $\rho$ between two scatterers. This distribution is characterized by its mean value $\langle \rho \rangle$ and its second moment $\langle \rho^2 \rangle$. The mean free path $\bar{\ell}=\langle \rho \rangle$ is inversely proportional to the density of scatterers, $N$, and to the scattering cross section $\xi$, $\bar{\ell}=1/(N\xi)$ \cite{Rogers2008}.

So far very few systems involving light propagation have been reported to exhibit non-normal behavior. One of them is a random amplifying medium (RAM), constituted of amplifying fiber segments embedded in a passive scattering bulk \cite{Sharma2006}. The intensity distribution at the sample exit results from multiple scattering of the photons by the passive bulk as well as from successive passages through fiber segments, in which the photons are amplified. Therefore, the longer the segment, the stronger the amplification. The distribution of lengths of the fiber-segments is intentionally tailored so as to yield a L\'{e}vy-like intensity distribution of the RAM emission.  A second system, called ``L\'{e}vy glass" by its designers \cite{Barthelemy2008}, is an engineered solid material where the density of scattering medium (titanium dioxide nanoparticles) is modulated by non-scattering spheres of diameter-distribution tailored to yield a L\'{e}vy-like transmission spatial distribution. Both of these experimental observations of L\'{e}vy flights of photons \cite{Sharma2006,Barthelemy2008} are based on synthetic, engineered, \textit{spatial} inhomegeneities. As we shall see in the following sections, in spatially \textit{homogeneous} resonant atomic vapours, the distance a photon will travel between two scattering events depends not only on the scatterers density (constant) but also on the photon's frequency, so that the superdiffusive transport in these natural media originates in a \textit{spectral} inhomogeneity instead of a spatial one.

\subsection{Radiation trapping}
Radiation trapping is the name given to the phenomenon of resonant multi-scattering of light in atomic vapours. In such a process, incident photons are first absorbed by atoms in the gas medium because their frequency $\nu$ is close to that of an atomic transition, $\nu_0$. The absorption of the photons is stronger at the center of the resonance, i.e. at $\nu = \nu_0$, and, depending on the absorption spectral shape, decays more or less sharply on either side of the resonance ($\left|\nu -\nu_0\right| > 0$). The excited atoms eventually decay radiatively to the ground state and the emitted photon can be absorbed by another atom and so on. The absorption-emission process can occur at very high rates (typically a few 100 MHz at room temperature) in resonant atomic vapours, so that many of these processes may take place before the photons leave the cell, resulting in a time much longer than the mean atomic lifetime for the radiation to leave the vapour volume. Radiation trapping is thus a mechanism that needs to be accounted for in the understanding of light propagation in stellar media \cite{Boldyrev2003}, discharge lamps \cite{Rajaraman2004,Camparo2007}, gas- \cite{Bachurin1991},liquid- \cite{Hammond1980}  as well as solid-state \cite{Sumida1994,Eichhorn2009,Stoita2010} laser media, atomic line filters \cite{Molisch1999}, trapped cold atoms \cite{Hillenbrand1995,Balik2009}, collision \cite{Lintz1998,Molisch2000} and coherent \cite{Matsko2001,Matsko2002} processes in atomic vapours, optical pumping of alkali-metal vapours \cite{Rosenberry2007}.

The concern with radiation imprisonment goes back at least as far as the 1920s, when Compton \cite{Compton1922} and Milne \cite{Milne1926} described theoretically the diffusion of resonant light in absorbing media. Kenty \cite{Kenty1932} interpreted experimental measurements by Zemansky \cite{Zemansky1927,Zemansky1930} by taking into account the redistribution of frequency that takes place between absorption and reemission of light and ascertained that ``Abnormally long free paths are found to be of such importance as to enable resonance radiation to escape from a body of gas faster than has usually been supposed... It is found that, for a gas container of infinite size, [the diffusion coefficient, the average square free path, and the average free path] are all infinite". In other words, the diffusion model, based on the finiteness of the mean free path and second moment, does not rigorously apply to the problem of radiation trapping in resonant media. A few years later, Holstein proposed a description through an integro-differential equation \cite{Holstein1947}, which still constitutes the starting point of most formal descriptions of radiation trapping.  

\subsubsection{Photons absorption and transmission probabilities} \label{SecProb}
\begin{figure}[htb]
\centering
\includegraphics[width=6.0cm]{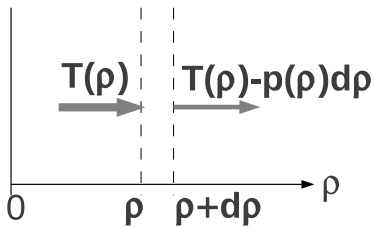}
\caption{Sketch of half space ($\rho > 0$) filled with scattering medium, to determine fraction of particles incident at $\rho$ and transmitted through a thin slice of thickness $d\rho$.}
\label{Fig1}
\end{figure}

Let us introduce a number of physical parameters useful to describe a single emission-absorption process in an isotropic medium. Let $T_{\nu}(\rho)$ be the probability that a photon of frequency $\nu$ travel a distance $\rho$ without being absorbed (transmission through $\rho$). $T_{\nu}(\rho)$ should have the limit values i) $T_{\nu}(0)$=1 and ii) $T_{\nu}(\infty)$=0. Let now $k(\nu)$ be the probability, per unit length, that the photon be absorbed. 
To determine the relationship between $k(\nu)$ and $T_{\nu}$, we write the transmission of the photon through a depth $\rho$ plus a thin slice of width $d\rho$ from ($\rho$) (see figure \ref{Fig1}) as the probability it arrives at $\rho$ times the probability it survives an additional distance $d\rho$: 

\begin{equation}
T_{\nu}(\rho+d\rho)= T_{\nu}(\rho)\left[1-k(\nu) d\rho\right].
\end{equation}

\noindent As, by definition, 

\begin{equation}
T_{\nu}(\rho+d\rho) - T_{\nu}(\rho) = \frac{\partial T_{\nu}}{\partial \rho} d\rho,
\label{partialT}
\end{equation}

\noindent then

\begin{equation}
T_{\nu}(\rho)= e^{-k(\nu)\rho}.
\label{Trans}
\end{equation}
 
\noindent This is the Beer-Lambert law \cite{Beerlaw}, verifying assumptions i) and ii) above.

The step-size distribution for a given frequency $\nu$ is $p_{\nu}(\rho)$. It means that of all the $T_{\nu}(\rho)$ photons of frequency $\nu$ created at position $\rho=0$ that have arrived at position $\rho$, $p_{\nu}(\rho) d\rho$ will not survive an additional slice of depth $d\rho$, so that $p_{\nu}(\rho) d\rho$ is the probability that these photons be absorbed between $\rho$ and $\rho + d\rho$. Again, we write the transmission of the photons through a depth $\rho$ plus a thin slice of width $d\rho$ from $\rho$ (see figure \ref{Fig1}):

\begin{equation}
T_{\nu}(\rho+d\rho) = T_{\nu}(\rho)-p_{\nu}(\rho)d\rho.
\label{TransAbs}
\end{equation}

\noindent From Eqs. \ref{partialT}, \ref{Trans} and \ref{TransAbs}, it follows that 

\begin{equation}
p_{\nu}(\rho)= -\frac{\partial T_{\nu}(\rho)}{\partial \rho}= k(\nu)e^{-k(\nu)\rho}.
\end{equation}

\noindent This expression describes the single-path length distribution for photons of frequency $\nu$ propagating in a scattering medium characterized by the absorption spectrum $k(\nu)$. 
We can determine the frequency-dependent mean free path $\bar{\ell}_{\nu}$ for the photons in the scattering medium as being the mean value of the step-length distribution:

\begin{equation}
\bar{\ell}_{\nu} = \int_{0}^{+\infty}{\rho \; p_{\nu}(\rho)d\rho} = \int_{0}^{+\infty}{\rho \;k(\nu) e^{-k(\nu) \rho}}=\frac{1}{k(\nu)}. 
\label{ellnu}
\end{equation}

\noindent Equation (\ref{ellnu}) tells us that we can tune the mean free path of a monochromatic beam of photons at frequency $\nu$ simply by varying the absorption coefficient $k(\nu)$ at this frequency. It is intuitively reasonable, as we may see in section \ref{specshapes}, that $k(\nu)$ is proportional to the atomic density, so that tuning of $\bar{\ell}_{\nu}$ can be achieved through control of the atomic density. If the light incident on the scattering medium is not monochromatic but has instead a spectral distribution  $\Theta(\nu)$, the resulting path-length distribution corresponds to an averaging of $p_{\nu}(\rho)$ over the (emission) spectrum $\Theta(\nu)$:

\begin{equation}
p(\rho) = \int_{-\infty}^{+\infty}{\Theta(\nu) \; p_{\nu}(\rho) \; d\nu}=\int_{-\infty}^{+\infty}{\Theta(\nu)\; k(\nu)\;e^{-k(\nu)\rho}\;d\nu}.
\label{single}
\end{equation} 

\noindent The moments of this distribution are given by:

\begin{eqnarray}
\langle \rho^q\rangle &=&\int_{0}^{+\infty}{\rho^q \; p(\rho) \; d\rho} \nonumber \\
&=&\int_{-\infty}^{+\infty}{d\nu \; \Theta(\nu) \; k(\nu) \int_{0}^{+\infty}{\rho^q e^{-k(\nu)\rho}d\rho}} \nonumber \\
&=&\int_{-\infty}^{+\infty}d\nu \; \Theta(\nu) \; k(\nu) \; \frac{\Gamma(q+1)}{k^{q+1}(\nu)} \nonumber \\
&=&q!\int_{-\infty}^{+\infty}d\nu \; \frac{\Theta(\nu)}{k^q(\nu)},
\label{momentq}
\end{eqnarray}

\noindent with $\Gamma$ the gamma function. In the special case of an incident laser beam, considered as monochromatic with frequency $\nu_L$, the photons' spectral distribution is $\Theta (\nu)= \delta (\nu - \nu_L)$ and the moments of the photons' steps distribution, given by

\begin{equation}
\langle \rho^q\rangle = \frac{q!}{k^q(\nu_L)},
\end{equation}  

\noindent are finite, as long as $k(\nu_L) \neq 0$. Particularly, for $q=1$, we again find the mean free path of equation (\ref{ellnu}). Similarly, in a vapour illuminated by any far-from-resonance spectral distribution ($\Theta (\nu)= W_{\infty}(\nu)$), the absorption spectral distribution can be considered constant ($k(\nu) \approx k_\infty$) so that, once again, the moments are finite and the diffusion of light is normal, as it is in any non-resonant, homogeneous scattering medium such as fog or diluted milk:

\begin{equation}
\langle \rho^q\rangle = \frac{q!}{k_\infty^q} \; \int_{-\infty}^{+\infty}{d\nu \; W_{\infty}(\nu)}.
\end{equation} 

\noindent On the other hand, the moments given by equation (\ref{momentq}) can be shown to diverge for all the spectral lineshapes $\Theta(\nu)$ and $k(\nu)$ usually associated with atomic transitions, specifically the Doppler, Lorentzian and Voigt lineshapes. It means that the random walk in an atomic vapour for photons with any of these spectral distributions cannot be characterized by a diffusion coefficient ($\propto \langle \rho^2\rangle$) and, in some cases, not even by a mean free path $\langle \rho\rangle$, thus qualifying propagation of photons in a resonant vapour as anomalous diffusion. 

\subsubsection{Frequency redistribution}
Clearly therefore, the relationship between the emission and the absorption spectra is a key element of the diffusion regime of photons in the vapour. This relation may evolve in the medium, due to multi-scattering, and is regulated by processes of \textit{frequency redistribution}. In optically thin samples of cold atoms, neither atomic collisions nor residual atomic motion (Doppler broadening smaller than the natural line width) are sufficient to significantly change the frequency of scattered photons before these escape the sample, so that no frequency redistribution takes place and the emission spectrum remains unchanged (monochromatic laser beam for example), distinct from the atomic absorption spectrum. Partial frequency redistribution (PFR) can be observed in stellar atmospheres and plasmas \cite{Mihalas1978,Voslamber1978} and in optically thick cold atomic samples, where the accumulation of tiny frequency redistribution due to residual motion of the atoms can become significant due either to a small mean free path (resonant light) or to large sample dimensions \cite{Pierrat2009}. In the complete frequency redistribution ͑(CFR͒) case, the frequencies of the scattered photons are fully decorrelated from the frequencies of the incident ones, for instance in situations where a high collision rate $\gamma_{col}$ in the vapour destroys correlations between photon absorption and ͑re͒emission events, i.e. where $\gamma_{col}\times \tau \gg 1$, with $\tau$ the lifetime of the excited state (CFR in the atom´s frame). In near-resonant thermal vapours, the frequency of the emitted photon is statistically independent of the frequency of the absorbed one through averaging over the Maxwellian velocity distribution (due to the Doppler effect). This (multiple) resonant scattering process very efficiently redistributes the photons' frequencies, leading to CFR in the laboratory frame and to coincidence of the emission and absorption spectra ($\Phi(\nu)= k(\nu)$). The single-step length distribution (equation (\ref{single})) is thus:

\begin{equation}
p(\rho) = \int_{-\infty}^{+\infty}{k^2(\nu)e^{-k(\nu)\rho}d\nu}
\end{equation} 

\noindent and the mean free path of the photons in the vapour becomes

\begin{equation}
\bar{\ell} = \int_{0}^{+\infty}{\rho \; p(\rho) \; d\rho}. 
\end{equation}

\section{L\'{e}vy flights in resonant atomic vapours}
As said above, the propagation of light in a scattering medium can be described as a random walk of photons within the medium. When the medium is near-resonant, the propagation of light is perturbed by the radiation trapping process: a photon incident in the medium (a laser photon for example) can be repeatedly absorbed and re-emitted in the medium. The multiple scattering process is now equivalent to a random walk in real space with frequency-dependent steps length, so that the spectral distribution of the photons, aided by the frequency redistribution process, is the key element determining their diffusion regime. 
In the case of Complete Frequency Redistribution, the frequency of the emitted photon is totally independent of the frequency of the absorbed one and is exclusively determined by the transition's spectral shape, be it Gaussian, Lorentzian or Voigt. As aforementioned in section \ref{SecProb} and as we will show in the next section, the single-step length distribution $p(\rho)$ decays asymptotically as $p(\rho)\sim 1/\rho^{(\alpha)}$ ($\rho \longrightarrow \infty$) more slowly than $1/\rho^3$. This slow decay characterizes broad (long tail) steps distributions, for which all moments $\langle\rho^q\rangle$ with $q\geq(\alpha-1)$ diverge. Thus, as first signalised by Kenty \cite{Kenty1932} and Holstein \cite{Holstein1947}, the assumption of a mean free path for the photons is not fulfilled in resonant vapours and a diffusion-like description of radiation trapping is not adequate. The relation with long-tailed distributions, studied in the 1930s by L\'{e}vy \cite{Levy}, and the theoretical labeling of incoherent radiation trapping as more a system exhibiting superdiffusive behavior, were made by Pereira \textit{et al.} \cite{Pereira2004} and further studied \cite{Berberan2006} in the early 2000's. Experimental confirmation followed a few years later \cite{Mercadier2009}.

\subsection{Spectral shapes of photons scattering}
\label{specshapes}
We may now focus on the spectral shapes more particularly associated with resonant light-atom interactions. They are given by :

\begin{equation}
k(\nu)= N \xi_0 f(\nu),
\end{equation} 

\noindent with $N$ the atomic density, $\xi_0$ the resonant cross-section for atoms at rest and $f(\nu)$ the normalized spectral shape, specific of the light-atom interaction regime. $\xi_0= \lambda^ 2/2\pi g_1/g_2$, where $\lambda$ is the wavelength and $g_1/g_2$ the ratio of degeneracy factors. 

The homogeneous Lorentz line shape characterizes systems with homogeneous broadening, due, for example, to spontaneous radiative decay or collisions, and is given by:

\begin{equation}
f_L(\nu)= \frac{\Gamma_N^2 /4}{4 \pi^ 2 \delta^ 2+\Gamma_N^2 /4},
\label{Lorentz}
\end{equation}

\noindent with $\delta=\nu-\nu_0$, the photon's detuning in relation to the transition's frequency $\nu_0$. $f_L(\nu)$ has a full width at half maximum (FWHM) $\Gamma_N/2\pi$ and an asymptotical decay 

\begin{align}
p_L(\rho)&\sim \frac{1}{\rho^{3/2}}, \\
&\rho \longrightarrow \infty \nonumber
\end{align}

\noindent i.e. with $\alpha=3/2$. As mentioned in section \ref{SecProb}, this characterizes a broad distribution with all moments diverging, i.e. not even a mean free path can be determined for the photon leaps.

The Doppler line shape, reflecting the velocity distribution of Doppler shifts, is given by:

\begin{equation}
f_D(\nu)= \frac{\Gamma_N \lambda}{4\sqrt{\pi}v_0} e^{x^2},
\label{Doppler}
\end{equation}

\noindent with the reduced frequency $x=\frac{\delta}{\nu_0} \frac{c}{v_0}$. $v_0=\sqrt{\frac{2 k_B T}{m}}$ is the most probable speed of the atoms of mass $m$ at temperature $T$ and $k_B$ is Boltzmann's constant. The FWHM of $f_D(\nu)$ is $\Gamma_D/2\pi=2 v_0/\lambda \sqrt{\textrm{ln} 2}$.  

It can be shown \cite{Pereira2004} that:

\begin{align}
p_D(\rho)&\sim \frac{1}{r^2\sqrt{\textrm{ln}\rho}} \sim \frac{1}{\rho^2},\label{pDoppler} \\
&\rho \longrightarrow \infty \nonumber
\end{align}

\begin{figure}[htb]
\centering
\includegraphics[width=8.0cm]{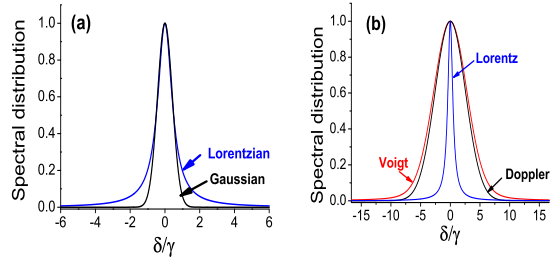}
\caption{Amplitude-normalized spectral distributions for two-level atoms. (a) Lorentzian (blue, equation (\ref{Lorentz})) and Gaussian (black, equation (\ref{Doppler})) shapes of same width (FWHM). (b) Lorentzian  (blue, width $\Gamma_N/2\pi$), Doppler (black, width $\Gamma_D/2\pi$) and resulting Voigt (red, equation (\ref{Voigt}), parameter $a$) lineshapes for natural linewidth $\Gamma_N$ and Doppler broadening $\Gamma_D=6 \Gamma_N$, i.e. Voigt parameter $a \approx 0.14$.}
\label{FigGLV}
\end{figure}

In figure \ref{FigGLV}(a) a Doppler (Gaussian) and a Lorentzian lineshape are drawn with same amplitude and width $\Gamma$ in order to make easier the comparison between their wings. The ``L\'{e}vy" behavior for the Lorentz spectral shape ($\alpha = 1.5$) is more dramatic than for the Doppler one ($\alpha = 2$) because the wings of the lorentzian distribution decrease more slowly than the Doppler ones and therefore go farther: the probability that a photon be re-emitted with a very large detuning and therefore travel a very long distance before being absorbed, is higher in the Lorentzian than in the Doppler case. 

The Voigt spectral distribution is a convolution of the two former ones:

\begin{eqnarray}
f_V(\nu)= \frac{\Gamma_N c}{4 \sqrt{\pi}v_0 \nu_0}\frac{a}{\pi}\int \frac{e^{-y^2} dy}{a^2+(x-y)^2},\label{Voigt} \\
a=\sqrt{\textrm{ln}2} \;\Gamma_N /\Gamma_D.\label{Voigta} 
\end{eqnarray}

\noindent with $a$ the Voigt parameter. It characterizes the interaction with light of atoms individually submitted to homogeneous, Lorentzian decaying processes (spontaneous emission, collisions,..) and moving around according to Gaussian Maxwell-Boltzmann velocity distribution.
As the wings of the Doppler lineshape tend very rapidly to zero, the asymptotic decay of the overall Voigt lineshape is the same as the Lorentzian one and the steps' distribution for a Voigt profile is also a broad distribution with $\alpha = 3/2$.

\begin{figure}
\centering
\includegraphics[width=8.0cm]{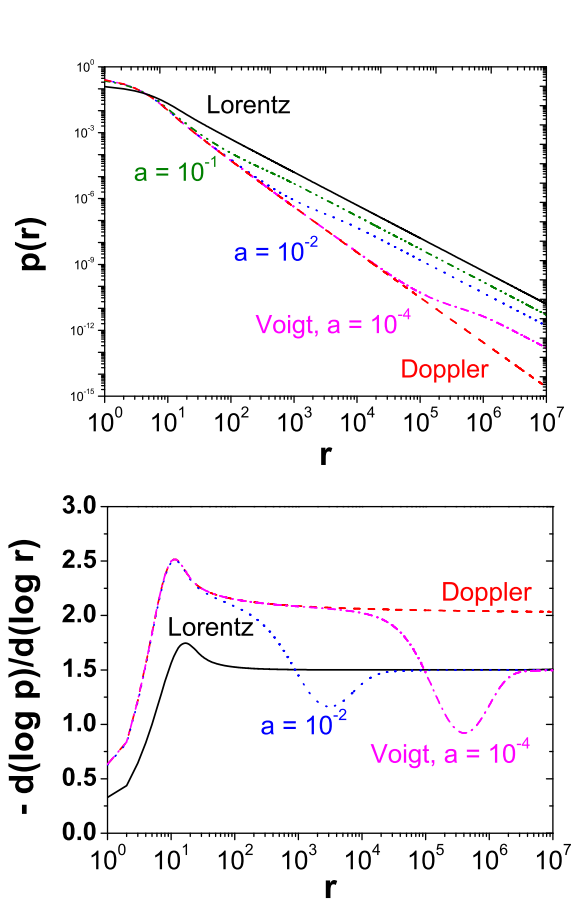}
\caption{Top: Step-length distribution for usual spectral lineshapes and CFR. Bottom: Derivative of step-length distributions. Solid black: Lorentz lineshape. Dash red: Doppler. Dash dot dot olive: Voigt, $a= 10^{-1}$. Dot blue: Voigt, $a= 10^{-2}$. Dot-dash magenta: Voigt, $a=10^{-4}$.}
\label{JumpSize}
\end{figure}

In figure \ref{FigGLV}(b) are shown the Lorentzian, Doppler and Voigt spectral distributions for a two-level atoms system of natural width $\Gamma_N$ (excited state lifetime $\tau= (\Gamma_N)^{-1}$) and Doppler width $\Gamma_D=6 \Gamma_N$. For example, this Doppler broadening would be the amount suffered by a vapour of rubidium atoms at a temperature of approximately 1.5 K \cite{Metcalf}. The resulting $a$ parameter of the Voigt distribution is thus of the order of 0.14. The $a$ parameter basically compares the natural and the Doppler line widths, $\Gamma_N$ and $\Gamma_D$, respectively (equation (\ref{Voigta})). In what is called the Doppler regime, the Doppler broadening of the lineshape is much larger than the homogeneous linewidth of the atomic transition and the lineshape tends to the Doppler one (small $a$). As the $a$ parameter increases, the Doppler broadening of the lineshape decreases in relation to the natural linewidth and the lineshape ultimately tends to the Lorentz one (cold atoms, for example). The configuration depicted in figure \ref{FigGLV}(b) corresponds to an intermediate case.  As $a$ increases from $a\ll 1$ to $a\geq 1$, the steps length distribution evolves from Doppler- to Lorentz-like, as can be seen in figure \ref{JumpSize}. The probability distributions in figure \ref{JumpSize} are actually graphed as functions of a dimensionless jump size, or opacity, $r = N \xi_0 \rho$. 

\section{Experimental characterization of L\'{e}vy flights of photons in an atomic vapour}
In cold atoms samples, the spectral lineshape is Lorentzian with natural linewidth (Doppler broadening $\Gamma_D \ll \Gamma_N$), but the scattering of photons is fundamentally elastic and frequency redistribution is insignificant, so that the hypothesis of L\'{e}vy flights of photons in such a medium does not apply to a first approximation \cite{Pierrat2009}. Up to now, the only atomic system in which photons L\'{e}vy flights have been directly measured (single-step length distribution with $\alpha\leq 3$) is a warm thermal vapour of rubidium atoms illuminated with a resonant laser \cite{Mercadier2009}. 

\begin{figure}
\centering
\includegraphics[width=8.0cm]{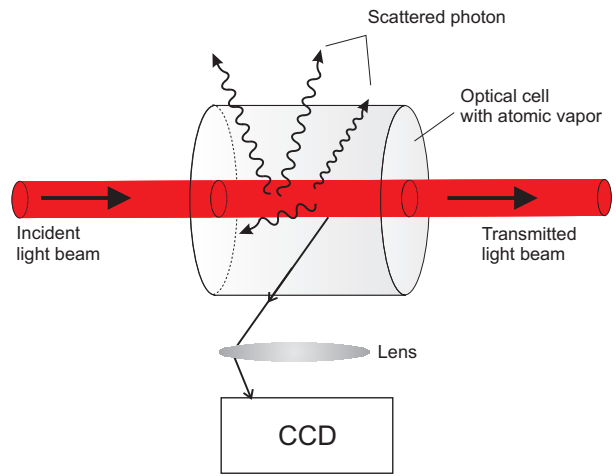}
\caption{Scheme of the set-up for experimental observation of distribution of step lengths for photons in a resonant atomic vapour.}
\label{Setup}
\end{figure}

The experimental set-up is shown schematically in figure \ref{Setup}. Photons with a spectral distribution $\Phi(\nu)$ are collimated and directed to a transparent cell containing an homogeneous atomic vapour of atoms. Photons scattered by the atoms and escaping the cell through a given solid angle are detected by a CCD camera, which forms images of the spatial profile of the fluorescence in the cell. Analysis of these images enables inference of the distribution of distances travelled by the incident photons in the scattering vapour.  

One major difficulty in this system, as in any multiple-scattering medium, is to infer the \textit{single-step} distribution from the observed result of multiple scattering, because multiple scattering erases the information of the photon's initial position. 

Let us detail a little the experimental aspects of this achievement as an illustration of the challenges faced when looking for signatures of microscopic processes in multiple scattering phenomena. Obviously, the higher the atomic density, the smaller the spatial scale allowing observation of the asymptotic slope (-$\alpha$) (see figure \ref{JumpSize}). It may then seem tempting to work with a very high atomic density, but this would increase the mean number of scattering events for the photons before they leave the cell and are detected, blurring  the  single-step contribution. The adopted compromise consists in slightly increasing the atomic density, so as to guarantee a reasonable rate of scattering events on a reasonable scale and to correct for the potentially perturbative multiple-scattering effects \cite{Mercadier2009}. The order of magnitude of the resonant cross-section for alkali atoms is a few $10^{-14}$ m$^2$ \cite{Metcalf}. Their natural linewidth is $\Gamma_N / 2 \pi \approx   6 \times 10^6$ s$^{-1}$. In a warm vapour ($T$ between 300 and 350 K), the Doppler broadening of the atomic spectral response is $\Gamma_D / 2\pi \approx 380-2300 \times 10^6$ s$^{-1}$, i.e. the $a$ parameter (equation (\ref{Voigta})) falls in the range of $10^{-3}$ to $10^{-2}$ . In this case, the asymptotic regime would be reached for optical densities above $10^4$ (see figure \ref{JumpSize}), i.e. in the required conditions of relatively low density (a few $10^{16}$ atoms m$^{-3}$), would require observation lengths larger than a few tens or hundreds of centimeters. The spatial scale available for experimental investigation is typically a few centimeters (let us say, up to 10 cm, the length of a quite long optical cell),  so what can actually be measured is the local slope $A$ (see figure \ref{JumpSize}, bottom), which tends to $-\alpha$ with increasing optical density. 

\begin{figure}
\centering
\includegraphics[width=8.0cm]{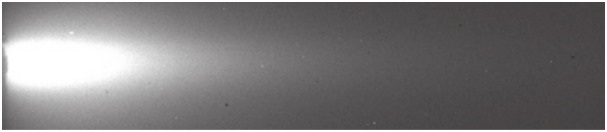}
\caption{Image of the fluorescence in the optical cell for incident photons having a Doppler-broadened  spectral distribution. (Reproduced from \cite{Mercadier2009}.)}
\label{Fluo}
\end{figure}

\begin{figure}
\centering
\includegraphics[width=8.0cm]{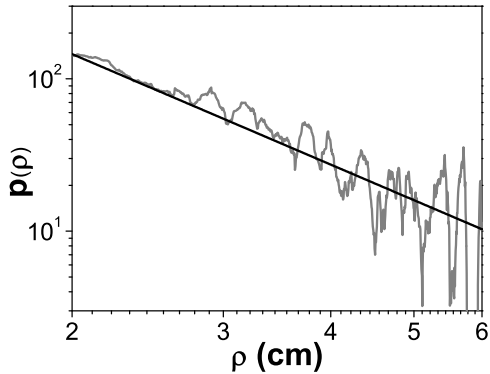}
\caption{Dark gray line: Jump-length distribution on the cell´s axis, as determined from the spatial profile of fluorescence. Black line: fit by power law $p(\rho) \propto \rho^{-\alpha}$, with $\alpha = 2.41 \pm 0.12$ (see \cite{Mercadier2009}) }
\label{Exp}
\end{figure}
Figure \ref{Fluo} shows an image of the fluorescence in the 7cm-long optical cell (see figure \ref{Setup}), as detected by the CCD camera. A bright point in the image indicates the position where a photon has been scattered towards the collection optics and the CCD camera. The brighter the spot, the higher the fluorescence intensity at the corresponding position. A thin slice of the image, a few tens of pixels wide along the cell's axis, is selected: it corresponds to the volume directly illuminated by the incident photons and where all the first-order scattering events take place (first scattering of an incident photon by an atom in the vapour). Above and below the central slice, the photons have necessarily been scattered at least once in the vapor and contribute to the multiply scattered signal. In figure \ref{Exp} the intensity in the central slice is plotted as a function of the position $x$ along the cell's axis, after correction for the multiple scattering contribution (see \cite{Mercadier2009}). The profile is plotted in log-log scale, in which a power law is displayed as a straight line. It is here well fitted by a power law $p(\rho) \propto \rho^{-\alpha}$, with $\alpha = 2.41 \pm 0.12$ \cite{Mercadier2009}.
In any case, the local slope $|A|$ is always smaller than 3, i.e. no  normal diffusion model can describe the propagation of light in such a medium. Another difficulty arises if the CFR condition is to be fulfilled (if it is not, complex frequency redistribution functions have to be determined \cite{AlvesPereira2007}). The frequency distribution of the impinging photons should therefore be the same as the absorption spectrum of the atoms in the target cell, i.e. a Voigt one with same $a$ parameter as the vapour. While this condition is not completely fulfilled in \cite{Mercadier2009}, an ingenious set of configurations (basically, the incident photons are prepared so as to result from a growing number of frequency-redistributing scattering events) allows the observation of the thermalisation of light, i.e. of the evolution from partially to almost completely frequency-redistributed incident photons, with a measured coefficient $\alpha$ always $\alpha \leq 3$, decreasing and tending to the expected value of $\approx 2$ \cite{Mercadier2009,Chevrollier2010}.    
      
\section{Conclusion}
We have shown how the propagation of light in resonant atomic vapours can exhibit a superdiffusive behavior, characterized by rare, long jumps between two scattering events. When present, these ``L\'{e}vy flights" dominate the dynamics of the system and may obliterate long sequences of apparently normal, Gaussian-like behavior (Gaussian distribution of jump lengths around a mean value). The physical origin of these rare events is the fact that, due to frequency redistribution, a photon have a very small but finite probability of being scattered with a frequency far from resonance, i.e. of flying a very long distance in the medium, turned almost transparent for this photon. The non-negligible probability of these long jumps ``lifts" the tail of the jump-size distribution: such ``long-tail" distributions decay asymptotically as power laws, characteristic of scale-free phenomena. L\'{e}vy flights occur for distributions $p(x)$ decaying slower than $1/x^3$, for which the second moment $\langle x^2 \rangle$, key element of normal statistics, diverges. L\'{e}vy flights have been observed in many other systems but emphasis has been placed on the atomic physics point of view, in which resonant atomic vapours represent the basic object of research. It is interesting that such a simple system, a paradigm of the most basic light-matter interaction mechanisms, still stirs curiosity and holds surprises. Atomic vapours actually represent small, tunable table-top systems where, depending on the conditions (temperature, atomic density), different diffusion regimes can be studied, as well as transitions between them. The limited spatial extension of experimental setups actually imposes a cutoff on the jump-length distribution. It has been shown \cite{Mantegna1994} that, although such a \textit{truncated L\'{e}vy flight} has finite variance, it may take a huge number of individual events for the resulting multi-scattering process to converge to a Gaussian one. The measurement of the segment of the distribution of individual jump lengths that is experimentally accessible is not affected by this cutoff restriction and the number of scattering events in a typical atomic vapour cell or lamp is in general not sufficient to characterize the diffusion regime as a normal one. It may however affect the actual diffusion regime in \textit{astrophysical samples}.
However restricted the example treated in this article may seem, it illustrates a subject belonging to the much more general topics of nonlinear dynamics and non-normal statistics, at the frontier of current knowledge. The advances in theoretical tools and experimental techniques allow us to extend our comprehension of the physical world, leading to the exploration of qualitatively new concepts. These are likely to lead to improved control over the systems in question.  

\label{lastpage}


\begin{thebibliography}{99}
\bibitem{Einstein1905}Einstein A., 1905, \textit{Annalen der Physik}, \textbf{322}, 549.
\bibitem{Brockmann2003}Brockmann D. and Geisel T., 2003, \textit{Phys. Rev. Lett.}, \textbf{90}, 170601. 
\bibitem{Levy}L\'{e}vy P., 1937, \textit{Th\'{e}orie de l'addition des variables al\'{e}atoires} (Gauthier-Villiers).
\bibitem{Stanley1996}Stanley H.E. \textit{et al.}, 1996, \textit{Physica A}, \textbf{224}, 302.
\bibitem{West1994}West B.J. and Deering W., 1994, \textit{Phys. Rep.}, \textbf{246}, 1.
\bibitem{Dieterich2008}Dieterich P., Klages R., Preuss R., and Schwab A., 2008, \textit{PNAS}, \textbf{105}, 459. 
\bibitem{Viswanathan1999}Viswanathan G.M., Buldyrev S.V., Havlin S., da Luz M.G.E., Raposo E.P. and Stanley H.E., 1999, \textit{Nature}, \textbf{401}, 911. 
\bibitem{Bartumeus2002}Bartumeus F., Catalan J., Fulco U.L., Lyra M.L., and Viswanathan G.M., 2002, \textit{Phys. Rev. Lett.}, \textbf{88}, 097901.
\bibitem{Nathan2002}Nathan R., Katul G.G., Horn H.S., Thomas S.M., Oren R., Avissar R.,
Pacala S.W., and Levin S.A., 2001, \textit{Nature}, \textbf{418}, 409.
\bibitem{Ramos2004}Ramos-Fernandez G., Mateos J.L., Miramontes O., Cocho G., Larralde H. and Ayala-Orozco B., 2004, \textit{Behav. Ecol. Sociobiol.}, \textbf{55}, 223.
\bibitem{Bartumeus2005}Bartumeus F., Da Luz M.G.E., Viswanathan G.M., and Catalan J., 2005, \textit{Ecology}, \textbf{86}, 3078. 
\bibitem{Viswanathan2008}Viswanathan G.M., Raposo E.P., da Luz M.G.E., 2008, \textit{Phys. Life Rev.}, \textbf{5}, 133. 
\bibitem{Humphries2010}Humphries N.E., Queiroz N., Dyer J.R.M., Pade N.G., Musyl M.K., Schaefer K.M., Fuller D.W., Brunnschweiler J.M., Doyle T.K., Houghton J.D.R., Hays G.C., Jones C.S., Noble L.R., Wearmouth V.J., Southall E.J. and Sims D.W, 2010, \textit{Nature}, \textbf{465}, 1066.
\bibitem{Viswanathan2011}Viswanathan G.M., da Luz M.G.E., Raposo E.P., and Stanley H.E., 2011, \textit{The Physics of Foraging}, (Cambridge: University Press).
\bibitem{Brockmann2006}Brockmann D., Hufnagel L. and Geisel T., 2006, \textit{Nature}, \textbf{439}, 462.
\bibitem{Gonzalez2008}Gonz\'{a}lez M.C., Hidalgo C.A. and Barab\'{a}si A.-L., 2008, \textit{Nature}, \textbf{453}, 779.
\bibitem{Stollenwerk2009}Stollenwerk N. and Boto J.P., 2009, \textit{AIP Conf. Proc.}, \textbf{1168}, 1548.
\bibitem{Mandelbrot1960}Mandelbrot B., 1960, \textit{International Economic Review}, \textbf{1}, 79.
\bibitem{Solomon1993}Solomon T.H., Weeks E.R., Swinney H.L., 1993, \textit{Phys. Rev. Lett.}, \textbf{71}, 3975.
\bibitem{Bardou1994}Bardou F., Bouchaud J.P., Emile O., Aspect A., and Cohen-Tannoudji C., 1994, \textit{Phys. Rev. Lett.}, \textbf{72}, 203.  
\bibitem{Venkataramani1997}Venkataramani S.C, Antonsen T.M., Jr., and Ott E., 1997, \textit{Phys. Rev. Lett.}, \textbf{78}, 3864.
\bibitem{Shimizu2001}Shimizu K.T., Neuhauser R.G, Leatherdale C.A., Empedocles S.A., Woo W.K., and Bawendi M.G., 2001, \textit{Phys. Rev. B}, \textbf{63}, 205316.
\bibitem{Tamura2002}Tamura K., Hidaka Y., Yusuf Y., and Kai S. , 2002, \textit{Physica A}, \textbf{306}, 157. 
\bibitem{Drysdale2004}Drysdale P.M. and Robinson P.A., 2004, \textit{Phys. Rev. E}, \textbf{70}, 056112.
\bibitem{Sharma2006}Sharma D., Ramachandran H., and Kumar N., 2006, \textit{Opt. Letters}, \textbf{31}, 1806. 
\bibitem{Barthelemy2008}Barthelemy P., Bertolotti J. and Wiersma D., 2008, \textit{Nature}, \textbf{453}, 495.
\bibitem{Boldyrev2003}Boldyrev S. and Gwinn C.R., 2003,  \textit{Phys. Rev. Lett.}, \textbf{91}, 131101.
\bibitem{Klafter1996}Klafter J. Shlesinger M.F., and Zumofen G., 1996, \textit{Phys. Today}, \textbf{49}, 33.
\bibitem{Kenty1932}Kenty C., 1932, \textit{Phys. Rev.}, \textbf{42}, 823.
\bibitem{Rogers2008}Rogers G.L., 2008, \textit{J. Opt. Soc. Am. A}, \textbf{25}, 2879.
\bibitem{Rajaraman2004}Rajaraman K. and Kushner M.J., 2004, \textit{J. Phys. D: Appl. Phys.}, \textbf{37}, 1780.
\bibitem{Camparo2007}Camparo J.C. and Mackay R., 2007, \textit{J. Appl. Phys.}, \textbf{101}, 053303.
\bibitem{Bachurin1991}Bachurin B.A., 1991, \textit{Sov. J. Quantum Electron.}, \textbf{21}, 42.
\bibitem{Hammond1980}Hammond P.R. and Nelson R., 1980, \textit{IEEE J. Quantum Electron.}, \textbf{QE-16}, 1161.
\bibitem{Sumida1994}Sumida D.S. and Fan T.Y., 1994, \textit{Opt. Lett.}, \textbf{19}, 1343.
\bibitem{Eichhorn2009}Eichhorn M., 2009, \textit{Appl. Phys. B}, \textbf{96}, 369.
\bibitem{Stoita2010}Stoita A., Vautey T., Jacquier B. and Guy S., 2010, \textit{J. Luminescence}, \textbf{130}, 1119.
\bibitem{Molisch1999}Molisch A.F. and Oehry B.P., 1999, \textit{Radiation Trapping in Atomic Vapours}, (Oxford: Oxford University Press).
\bibitem{Hillenbrand1995}Hillenbrand G., Burnett K. and Foot C.J., 1995, \textit{Phys. Rev. A}, \textbf{52}, 4763.
\bibitem{Balik2009}Balik S., Havey M.D., I.M. Sokolov I.M. and Kupriyanov D.V., 2009, \textit{Phys. Rev. A}, \textbf{79}, 033418.
\bibitem{Lintz1998}Lintz M. and Bouchiat M.A., 1998, \textit{Phys. Rev. Lett.}, \textbf{80}, 2570.
\bibitem{Molisch2000}Molisch A.F., Oehry B.P. and Magerl G., 2000, \textit{Physica Scripta}, \textbf{T86}, 55.
\bibitem{Matsko2001}Matsko A.B., Novikova I., Scully M.O. and Welch G.R., 2001, \textit{Phys. Rev. Lett.}, \textbf{87}, 133601. 
\bibitem{Matsko2002}Matsko A.B., Novikova I., and Welch G.R., 2002, \textit{J. Mod. Optics}, \textbf{49}, 367. 
\bibitem{Rosenberry2007}Rosenberry M.A., Reyes J.P., Tupa D. and Gay T.J., 2007, \textit{Phys. Rev. A}, \textbf{75}, 023401, and ref. therein.
\bibitem{Compton1922}Compton K.T., 1922, \textit{Phys. Rev.}, \textbf{20}, 283. 
\bibitem{Milne1926} Milne E.A., 1926, \textit{J. Lond. Math. Soc.}, \textbf{1}, 40. 
\bibitem{Zemansky1927}Zemansky M.W., 1927, \textit{Phys. Rev.}, \textbf{29}, 513. 
\bibitem{Zemansky1930}Zemansky M.W., 1930, \textit{Phys. Rev.}, \textbf{36}, 919. 
\bibitem{Holstein1947}Holstein T., 1947, \textit{Phys. Rev.}, \textbf{72}, 1212.
\bibitem{Beerlaw}Grynberg G., Aspect A., Fabre C., 2010, \textit{Introduction to Quantum Optics: From the Semi-classical Approach to Quantized Light}, (Cambridge: University Press).
\bibitem{Mihalas1978}Mihalas D., 1978, \textit{Stellar Atmospheres}, (San Francisco:Freeman, 2nd ed.).
\bibitem{Voslamber1978}Voslamber D. and Yelnik J.-B., 1978, \textit{Phys. Rev. Lett.}, \textbf{41}, 1233.
\bibitem{Pierrat2009}Pierrat R., Gr\'{e}maud B. and Delande D., 2009, \textit{Phys. Rev. A}, \textbf{80}, 013831.	
\bibitem{Pereira2004}Pereira E., Martinho J.M.G. and Berberan-Santos M.N., 2004, \textit{Phys. Rev. Lett.}, \textbf{93}, 120201.
\bibitem{Berberan2006}Berberan-Santos M.N., Pereira E.J.N., and Martinho J.M.G., 2006, \textit{J. Chem. Phys.}, \textbf{125}, 174308.
\bibitem{Mercadier2009}Mercadier N., Gu\'{e}rin W., Chevrollier M. and Kaiser R., 2009, \textit{Nature Physics}, \textbf{5}, 602.
\bibitem{Metcalf} Metcalf H.J. and van der Straten P., 1999, \textit{Laser Cooling and Trapping}, (New York: Springer-Verlag).
\bibitem{AlvesPereira2007}Alves-Pereira A.R., Nunes-Pereira E.J., Martinho J.M.G. and Berberan-Santos M.N., 2007, \textit{J. Chem. Phys.}, \textbf{126}, 154505.
\bibitem{Chevrollier2010}Chevrollier M., Mercadier N., Gu\'{e}rin W., and Kaiser R., 2010, \textit{Eur. Phys. J. D}, \textbf{58}, 161.
\bibitem{Mantegna1994}Mantegna R.N. and Stanley H.E., 1994, \textit{Phys. Rev. Lett.}, \textbf{73}, 2946.

\end{thebibliography}
\end{document}